\documentclass[proceedings, preprint]{rmaa}

\usepackage{paralist}

\usepackage{psfrag,color}

\font\sevenrm=cmr7
\newcommand\uchiie{\hbox{UC~H~{\sevenrm II}~+~EE~}}
\newcommand\uchii{\hbox{UC~H~{\sevenrm II}~}}
\newcommand\uchiis{\hbox{UC~H~{\sevenrm II}~s~}}
\newcommand\hii{\hbox{H{ }{\sevenrm II}{ }}}

\newcommand\degree  {\ifmmode{^\circ}\else{$^{\circ}$}\fi}
\newcommand\amin   {\ifmmode{^\prime}\else{$^\prime$}\fi}
\newcommand\pamin  {\ifmmode{{\rlap.}^\prime}\else{${\rlap.}^\prime$}\fi}
\newcommand\asec   {\ifmmode{^{\prime \prime}}\else{$^{\prime \prime}$}\fi}
\newcommand\pasec  {\ifmmode{{\rlap.}^{\prime \prime}}\else{${\rlap.}^{\prime \prime}$}\fi}

\newcommand\lax    {\ifmmode{_<\atop^{\sim}}\else{${_<\atop^{\sim}}$}\fi}
\newcommand\gax    {\ifmmode{_>\atop^{\sim}}\else{${_>\atop^{\sim}}$}\fi}

\def\arcsec {\hbox{$^{\prime\prime}$}}
\def\arcmin {\hbox{$^{\prime}$}}

\SetYear{2008}
\SetConfTitle{A Long Walk Through Astronomy: A Celebration of Luis Carrasco's 60th Birthday}

\title{\emph{Spitzer}--IRAC imagery and photometry of ultracompact H~II regions with extended emission\altaffilmark{1}} 

\author{
  E. de la Fuente,\altaffilmark{2}
  A. Porras,\altaffilmark{3}
  J. M. C. Grave,\altaffilmark{4}
  M. S. N. Kumar,\altaffilmark{4}
  M. A. Trinidad,\altaffilmark{5}
  S. Kurtz,\altaffilmark{6}
  S. Kemp, \altaffilmark{2}
  J. Franco,\altaffilmark{7}
  G. Quevedo\altaffilmark{2}}

\altaffiltext{1}{To be published in ''A Long Walk Through Astronomy: A Celebration of Luis Carrasco's 60th Birthday'' proceedings, RMxAA SC. Work based in de la Fuente Ph. D. Thesis, 2007, University of Guadalajara, M\'exico }

\altaffiltext{2}{Instituto de Astronom\'{\i}a y Metereolog\'{\i}a, Departamento de F\'{\i}sica, CUCEI, Universidad de Guadalajara. Av. Vallarta 2602, Col. Arcos Vallarta. C.P. 44130 Guadalajara, Jal. M\'exico (edfuente@astro.iam.udg.mx).}

\altaffiltext{3}{Instituto Nacional de Astrof\'{\i}sica, \'Optica y Electr\'onica (INAOE). Luis E. Erro Num. 1, Tonantzintla, Pue. M\'exico (aporras@inaoep.mx).}

\altaffiltext{4}{Centro de Astrof\'{\i}sica da Universidade do Porto. Rua das Estrelas, 4150-762 Porto, Portugal.}

\altaffiltext{5}{Departamento de Astronom\'{\i}a, Universidad de Guanajuato, P.O. Box 144, Guanajuato, GTO 36000, Mexico.}

\altaffiltext{6}{Centro de Radioastronom\'\i{}a y Astrof\'\i{}sica, Universidad Nacional Aut\'onoma de M\'exico, Apdo. Postal 3--72, 58089, Morelia, Mich. Mexico.}

\altaffiltext{7}{Instituto de Astronom\'{\i}a, Universidad Nacional Aut\'onoma de M\'exico, Apdo. Postal 70-264, 04510 M\'exico D.F., Mexico.}

\shortauthor{de la Fuente et al.}
\shorttitle{\emph{Spitzer}--IRAC of \uchiie}

\listofauthors{E. de la Fuente,  A. Porras,  J. M. C. Grave,  M. S. N. Kumar,  M. A. Trinidad,  S. Kurtz, S. Kemp, J. Franco, \& G. Quevedo}

\indexauthor{Porras, A.}
\indexauthor{de la Fuente, E.}
\indexauthor{Grave, J. M. C.}
\indexauthor{Kumar, M. S. N.}
\indexauthor{Trinidad, M. A.}
\indexauthor{Kurtz, S.}
\indexauthor{Kemp, S.}
\indexauthor{Franco, J.}
\indexauthor{Quevedo, G.}

\abstract{We present the results of a morphological study performed to a sample of 
Ultracompact (UC) H~II regions with Extended Emission (EE) using \emph{Spitzer}--IRAC 
imagery and 3.6~cm VLA conf. D radio-continuum (RC) maps. Some examples of the comparison between maps 
and images are presented. Usually there is an IR point source counterpart to 
the peak(s) of RC emission, at the position of the \uchii source. We find that the predominant EE morphology 
is the cometary, and in most cases is coincident with IR emission at 8.0~$\mu$m.
Preliminary results of \emph{Spitzer}--IRAC photometry of a sub-sample of 13 \uchii 
regions with EE (\uchiie) based on GLIMPSE legacy data are also presented. Besides, 
individual IRAC photometry was performed to 19 \uchii sources within these 13 regions. 
We show that \uchii sources lie on specific
locus, both in IRAC color-color and AM-product diagnostic diagrams. Counts of 
young stellar sources are presented for each region, and we conclude that a 
proportion of $\sim$2\%, $\sim$10\%, and $\sim$88\% of sources in \uchiie are, in average, 
Class~I, II, and III, respectively.}

\resumen{Presentamos los resultados de un estudio morfol\'ogico realizado a una muestra de regiones H~II ultracompactas (UC) con emisi\'on extendida (EE) usando im\'agenes \emph{Spitzer}--IRAC y mapas de radio-continnum (RC) del VLA a 3.6~cm en su configuraci\'on D. Presentamos algunos ejemplos de la comparaci\'on entre mapas e im\'agenes. Usualmente hay una fuente puntual infrarroja que es contraparte de el(los) pico(s) de emisi\'on en RC, en la posici\'on de la fuente \uchii. Encontramos que la morfolog\'{\i}a predominante de la EE es cometaria, y en la mayor\'{\i}a de los casos es coincidente con emisi\'on IR a 8.0~$\mu$m.
Tambi\'en presentamos resultados preliminares de la fotometr\'{\i}a \emph{Spitzer}--IRAC de una sub-muestra de 13 regiones \uchii con EE (\uchiie) basada en datos del legado de GLIMPSE. Adem\'as, realizamos fotometr\'{\i}a IRAC individual a 19 fuentes \uchii dentro de estas 13 regiones. Mostramos que las fuentes \uchii caen en lugares espec\'{\i}ficos, en ambos diagramas de diagn\'ostico, color-color de IRAC y del producto AM. Presentamos conteos de fuentes estelares j\'ovenes para cada regi\'on, y concluimos que una proporci\'on de $\sim$2\%, $\sim$10\%, y $\sim$88\% de las fuentes en \uchiie son, en promedio, Clase~I, II, y III, respectivamente.}

\addkeyword{H~II regions: Ultracompact}
\addkeyword{Stars: Pre-main sequence}
\addkeyword{Stars: High mass}

\begin{document}
\maketitle

\section{Introduction}
\label{sec:intro}

The ultracompact \hii (\uchii) regions (term coined by Israel, Habing \& de Jong 1973), are small (size $\leq$ 0.1 pc), dense ($\gtrsim$ 10$^4$ cm$^{-3}$), photoionized Hydrogen regions with high emission measures ($\geq$ 10$^7$ {${\rm pc\ cm}^{-6}$}), surrounding a recently formed ionizing (OB type) star(s). They are considered as good tracers of recent massive star formation, and generally they are surrounded by a natal dust `cocoon'.

Recent works show that several \uchii regions have extended emission (EE) at arcmin scales (de la Fuente 2007; F07 hereafter, de la Fuente et al. 2009; F09 hereafter, Ellingsen et al., 2005; Kim \& Koo, 2001; Kurtz et al. 1999). It is not clear if this EE is physically associated with the ultracompact (UC) emission at arcsec scales, but if it is, there will be strong implications and changes in our understanding of these objects, such as definition, environment, energetics and excitation mechanism (F07; F09; Kurtz 2002; Kurtz et al. 1999). This association can be performed comparing VLA with IR observations (F07; F09). The former traces ionized gas and the latter dust and environment at large scales, where previous VLA radio-observations only detect sources smaller than 20$''$--30$''$ (e.g. Kurtz, et al. 1994; Wood \& Churchwell 1989; using the conf. A and B), and extended emission at scales up to $\sim$~3$'$ (Kurtz et al. 1999) and to $\sim$~7--15$'$ (Condon et al. 1998; Kim \& Koo 2001) using the conf. D.

The \emph{Spitzer} legacy project GLIMPSE (Benjamin et al. 2003\footnote{see also \emph{http://www.astro.wisc.edu/glimpse/}}), via photometry and imaging, has provided excellent tools to study star formation regions, discover star clusters, and to characterize the YSO population (e.g. F07; F09; Chavarria et al. 2008; Kumar \& Grave 2007, KG07 hereafter; Hartmann et al. 2005 and references therein).

A summary of the results reported in de la Fuente Ph. D. thesis (2007), regarding the comparison between IRAC imagery and VLA low resolution maps is presented in \S~2 and \S~3, while the IRAC photometry analysis and young stellar classification also for \uchii sources, are presented in \S~4. Preliminary conclusions are given in \S~5.

\section{The de la Fuente 2007 Sample and Observations}
\label{sec:errors}

29 \uchii regions with EE were studied by F07. The selection criteria was to include regions with both: a) EE and suggested candidates according to Kurtz et al. (1999) and Kim \& Koo (2001), plus G35.20--1.74 (IRAS 18592+0108) and G19.60-0.23 (IRAS 18248--1158), and b) IR excess in the Wood \& Churchwell (1989) and Kurtz et al. (1994) samples (see F09).

The Radio Continuum (RC) maps were obtained from VLA conf. D observations at 3.6~cm. These observations trace the ionized gas and have resolutions $\sim$ 9$''$ and are sensitive up to $\sim$~3$'$ structures. In order to have VLA maps in conf. D at 3.6~cm from all sources, F07 combine several new observations with those presented in Kurtz et al. 1999. Nevertheless, although F07 made VLA multiresolution cleaning maps using conf. B, C, and D, here we will discuss only the conf. D maps.

22 of the 29 \uchii regions have infrared images available in all IRAC (Fazio et al. 2004) bands at 3.6, 4.5, 5.8, and 8.0~$\mu$m. They are listed in Table 1. Dust and PAH features (at 3.3, 6.2, 7.7, and 8.6~$\mu$m) can be detected in the 3.6, 5.8 and 8.0~$\mu$m bands. The 8.0~$\mu$m band is strongly dominated by PAH emission (predominant at 7.7~$\mu$m). Shocked emission is observed at 4.5~$\mu$m band (Cyganowski et al. 2008). The 3.6~$\mu$m band also shows the presence of stellar clusters and nebulosities (ionized or reflection nebula), and can be used as a tracer of photodissociation regions (PDRs) via PAH emission (e.g. Sloan et al. 1996 \footnote{\emph{http://isc.astro.cornell.edu/$\sim$sloan/library/1996/pahim9} \emph{6a.html}}).

\begin{table}[!t]\centering
\small
  \setlength{\tabnotewidth}{\columnwidth}
  \tablecols{4}
  \setlength{\tabcolsep}{\tabcolsep}
  \caption{Ultracompact \hii regions with IRAC Emission\tabnotemark{a}} \label{tab:irac}
  \begin{tabular}{lrrr}
    \toprule
    IRAS Source &  $\alpha$(2000) & $\delta$(2000) & Distance\tabnotemark{b}  \\
    \midrule
17559--2420 & 17 59 03 & --24 20 49 & 14.3\tabnotemark{1}\\
18060--2005 & 18 08 58 & --20 05 15 & 6.0\tabnotemark{2}     \\
18097--1825A & 18 12 40 & --18 24 21 & 13.5\tabnotemark{2}     \\
18222--1321 & 18 25 01 & --13 15 40 & 4.2\tabnotemark{3}      \\
18248--1158 & 18 27 38  & --11 56 42  & 3.5\tabnotemark{2}   \\
18319--0834 & 18 34 45  & --08 31 07  & 9.0\tabnotemark{1}  \\
18311--0809 & 18 33 53  & --08 07 26  & 8.9\tabnotemark{2}     \\
18353--0628 & 18 38 03  & --06 23 47 & 9.3\tabnotemark{2}      \\
18402--0417 & 18 42 58  & --04 14 05  & 9.1\tabnotemark{3}     \\
18469--0132 & 18 49 32 & --01 29 04  & 8.9\tabnotemark{6}      \\
18496+0004 & 18 52 08   & +00 08 12  & 7.1\tabnotemark{4}    \\
18538+0216 & 18 56 24 & +02 20 38 & 3.6\tabnotemark{4}       \\  
18557+0358 & 19 00 16  & +04 03 13  & 9.9\tabnotemark{2}  \\
18593+0408 & 19 01 54   & +04 12 49  & 9.3\tabnotemark{4}     \\
19081+0903 & 19 10 35  & +09 08 31  & 11.7\tabnotemark{5}   \\
19110+1045 & 19 13 22  & +10 50 53 & 6.0\tabnotemark{4}    \\
19111+1048 & 19 13 28  & +10 53 37  & 6.0\tabnotemark{4}     \\
19120+0917 & 19 14 26  & +09 22 34 & 8.0\tabnotemark{4}    \\
19120+1103  & 19 14 21 & +11 09 04   & 6.0\tabnotemark{4}   \\
19181+1349  & 19 20 31 & +13 55 24 & 9.8\tabnotemark{4}       \\
19294+1836 & 19 31 43  & +18 42 52 & 7.9\tabnotemark{1}      \\
19442+2427 & 19 44 14  & +24 27 50  & 2.2\tabnotemark{4}   \\
    \bottomrule
    \tabnotetext{a}{From the F07 sample of 29 \uchii regions with EE}
    \tabnotetext{b}{Taked from: (1) Wood \& Churchwell (1989);(2) Churchwell et al. (1990); (3) Kurtz et al. (1994); (4) Araya et al. (2002); (5) Watson et al. (1997); (6) Kurtz et al. (1999).}

  \end{tabular}
\end{table}

\begin{figure*}[!t]
  \includegraphics[width=\columnwidth,height=7.5cm]{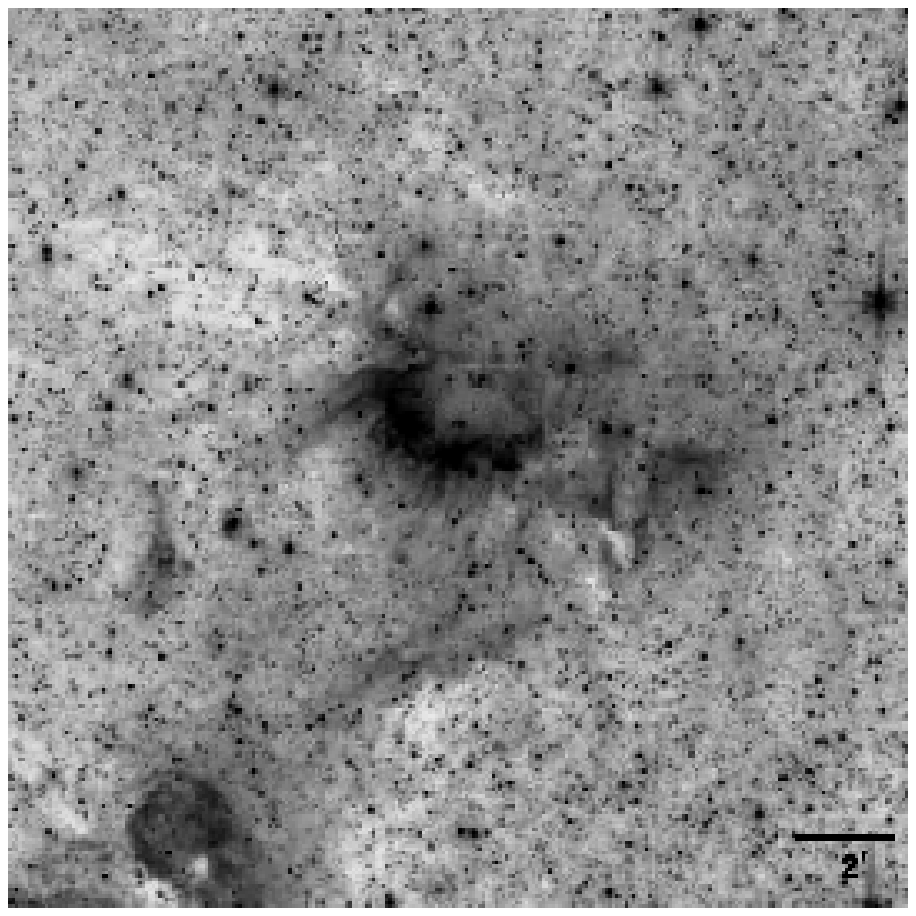}%
  \hspace*{\columnsep}%
  \includegraphics[width=\columnwidth,height=7.3cm]{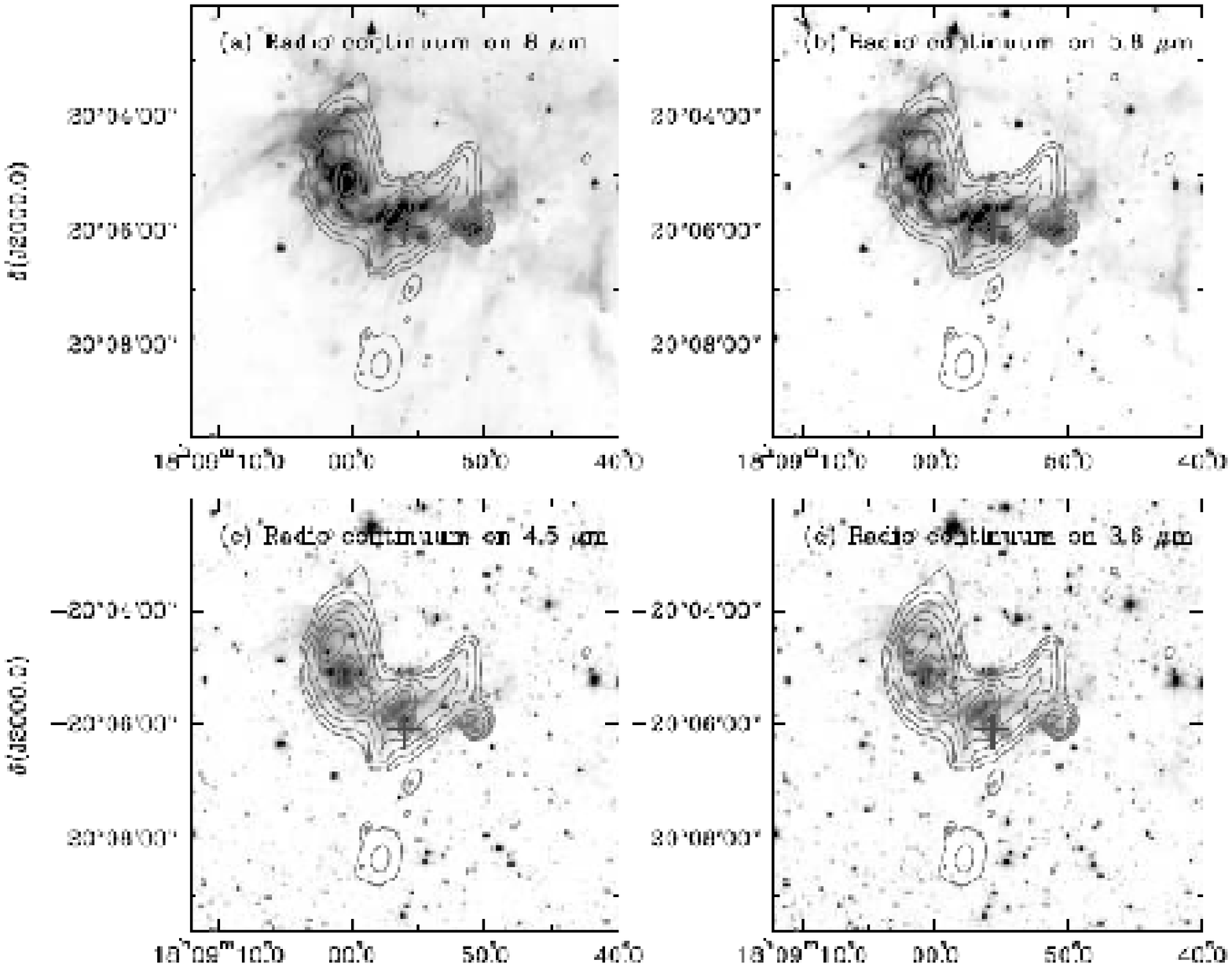}
  \caption{The \uchii region with EE IRAS 18060--2005 (G10.30--0.15). Left: Gray scale IRAC 4--bands combined image. North is up and East to the left. Right: IRAC images at a) 8.0~$\mu$m, b)5.8~$\mu$m, c)4.5~$\mu$m, and d) 3.6~$\mu$m with the VLA conf. D map (F07) superposed as contours (the EE). The crosses show the position of the \uchii region.}
  \label{fig:widefig1}
\vspace{0.5cm}
\end{figure*}

\begin{figure*}[!t]
  \includegraphics[width=\columnwidth,height=7.5cm]{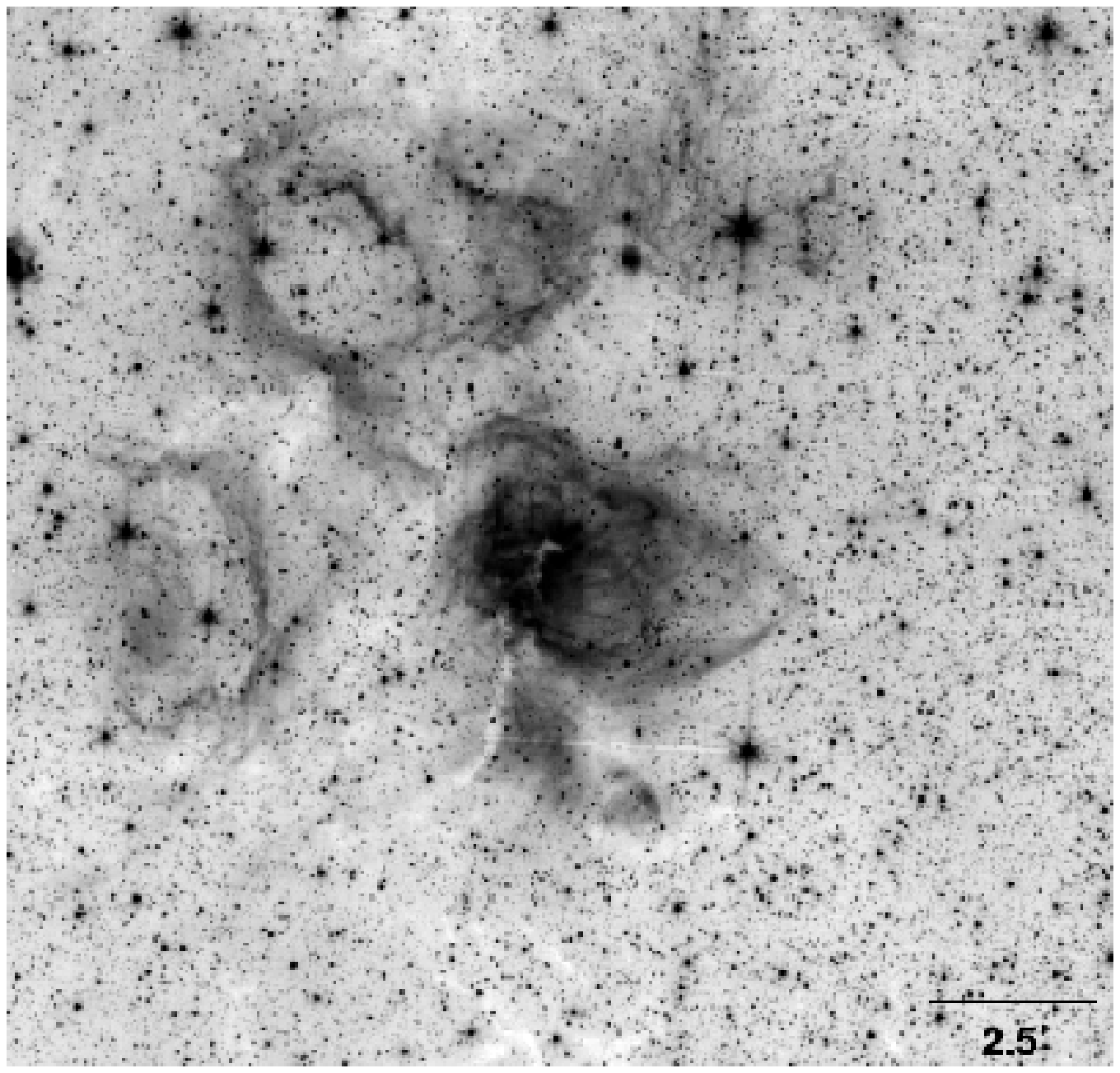}%
  \hspace*{\columnsep}%
  \includegraphics[width=\columnwidth,height=7.3cm]{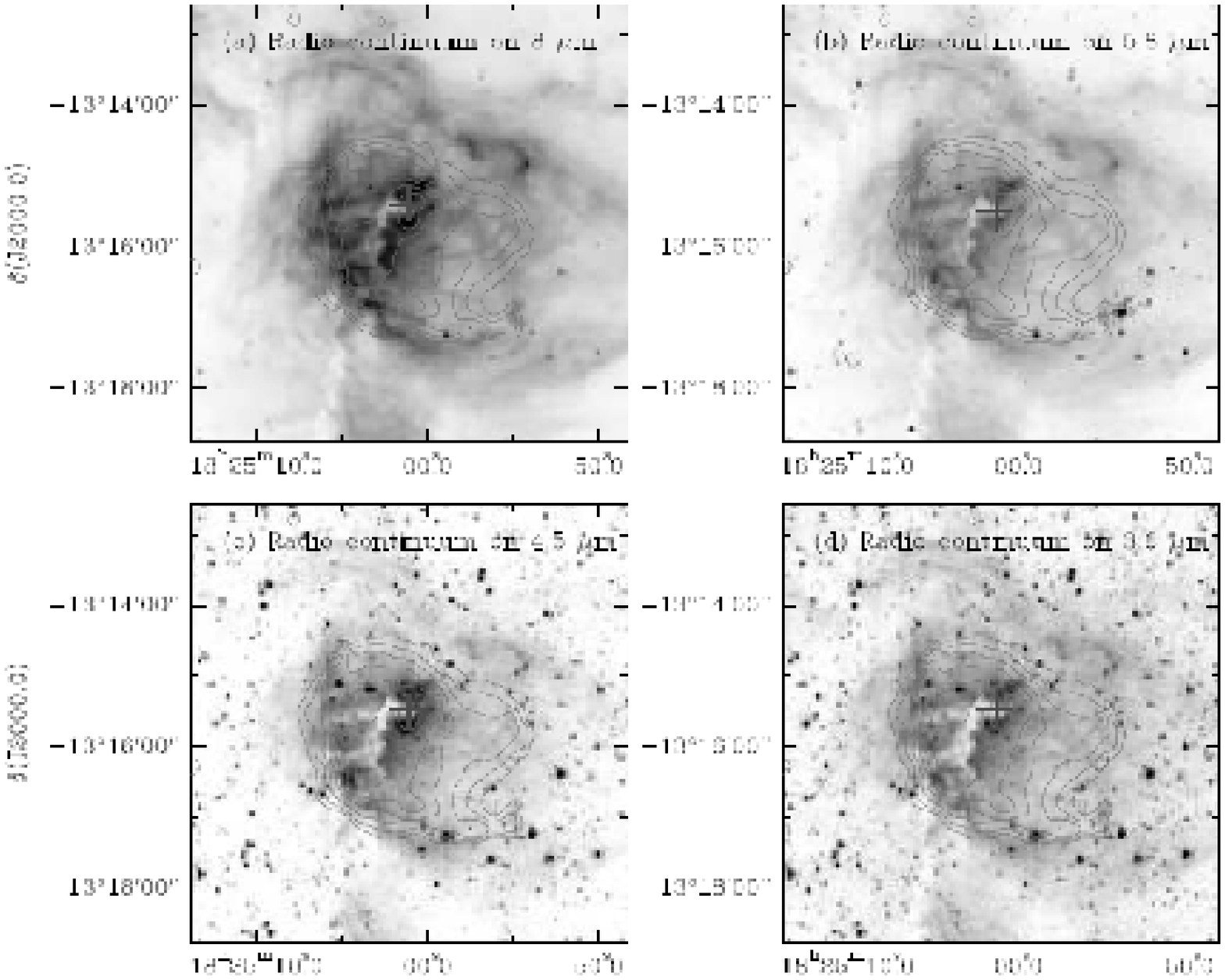}
  \caption{Same as Figure 1 for the \uchii region with EE IRAS 18222-1321 (G18.15--0.28)}
  \label{fig:widefig2}
\end{figure*}

\section{Maps and Images}
\label{sec:command}

Figures 1 and 2 show all IRAC images of the \uchii regions with EE IRAS 18060--2005 (G10.30--0.15) and IRAS 18222-1321 (G18.15--0.28), respectively, with a four-band gray scale image. Figure 3 shows the four-band combined IRAC image of IRAS 18577+0358 (G37.55--0.11). In these figures, the respective VLA conf. D maps are superposed as contours. 

The IR emission (IRE) has similar morphology in all bands suggesting that in these regions, both gas and dust coexist. As was expected, dust has a strong presence in these massive star formation regions (e.g., Wood \& Churchwell 1989; Kurtz et al. 1994), and plays a fundamental role in several physical processes related with \uchii regions with EE like their energetics (F07; F09). The IRE is brightest in the 8~$\mu$m band and decreases at lower wavelengths becoming sometimes undetectable in the 3.6~$\mu$m band. 

Coincidence between radio and IR morphologies is observed, confirming the presence of heated dust within the ionized gas. Morphologies are similar to those defined for \uchii regions (e.g., Wood \& Churchwell 1989: Kurtz et al. 1994): cometary (IRAS 18353--0628), core--halo (e.g. IRAS 19120+1103), bipolar (e.g. IRAS 18593+0408), and irregular or multipeaked. Nevertheless, the cometary morphology appears to dominate in our sample. The strongest IRAC emission is observed in the RC peaks, and in the cometary arcs. 

The PAH emission is a good tracer of the ``radiation temperature'' and the IRAC 8.0~$\mu$m band is dominated by this emission. The striking comparison between the IRAC images and the VLA RC maps (Figures 1 to 3) suggests that the EE is due to ionizing radiation.  However, soft UV radiation which may not significantly contribute to the overall \hii region could be an important ionization source for the EE.  In the 8.0~$\mu$m image, several knot--like sources are observed. They could be either star clusters or externally illuminated condensations.  Furthermore, the 8.0~$\mu$m band can effectively trace weak structures and has proven useful in unveiling the underlying physical structure of the dense core/cloud (e. g., Heitsch et al. 2007).

\begin{table*}[!t]\centering
  \setlength{\tabnotewidth}{0.9\textwidth}
  \tablecols{7}
  \setlength{\tabcolsep}{1.7\tabcolsep}
  \caption{Summary of {\emph Spitzer}--IRAC photometry of \uchii Sources\tabnotemark{a}} \label{tab:uchiis}
  \begin{tabular}{clrrccc}
    \toprule
   \uchii & \multicolumn{1}{c}{IRAS name} & \multicolumn{1}{c}{RA$_{2000}$} & \multicolumn{1}{c}{DEC$_{2000}$} & \multicolumn{1}{c}{[3.6]--[4.5]} & \multicolumn{1}{c}{[5.8]--[8.0]} & \multicolumn{1}{c}{$\alpha_{IRAC}$} \\
   Number & \multicolumn{1}{c}{} & \multicolumn{1}{c}{} & \multicolumn{1}{c}{} & \multicolumn{1}{c}{} & \multicolumn{1}{c}{} & \\
    \midrule
    1 & 18097--1825A& 273.16498 & --18.40563 &--0.04\,\,\,& 2.23 & 0.18 \\
    2 & 18222--1321 & 276.25469 & --13.26131 & 0.88 & 1.66  & 2.01 \\
    3 & 18248--1158 & 276.90905 & --11.94459 & 1.49 & 2.44  & 3.83  \\
    4 &             & 276.90557 & --11.94238 & 1.52 & 1.84  & 3.21  \\
    5 &             & 276.91150 & --11.94552 & 0.56 & 1.82  & 1.85  \\

    6 &             & 276.90965 & --11.94083 & 1.29 & 1.64  & 3.32  \\
    7 & 18311--0809 & 278.47304 & --8.12053  & 1.03 & 1.74  & 2.61  \\
    8 &             & 278.47186 & --8.11929  & 0.75 & 1.89  & 3.18  \\
    9 & 18353--0628 & 279.51210 & --6.39700  & 0.79 & 1.81  & 2.82  \\
   10 &             & 279.51308 & --6.40426  & 2.98 & 0.66  & 2.44  \\

   11 & 18402--0417 & 280.74260 & --4.23288  & 1.72 & 0.67  & 2.65  \\
   12 & 18469--0132 & 282.38774 & --1.48423  & 1.57 & 1.45  & 1.81  \\
   13 & 18538+0216  & 284.09413 &   2.34090  & 1.31 & 1.60  & 2.59  \\
   14 & 18577+0358  & 285.06682 &   4.05381  & 1.09 & 1.67  & 3.16  \\
   15 & 18593+0408  & 285.47342 &   4.21320  & 1.76 & 1.50  & 3.55  \\

   16 & 19120+0917  & 288.60926 &   9.37569  & 1.28 & 1.83  &  3.46  \\
   17 & 19120+1103  & 288.58885 &  11.15436  & 1.62 & 0.37  & -0.61  \\
   18 &             & 288.60677 &  11.15807  & 1.36 & 1.79  &  3.05  \\
   19 & 19181+1349  & 290.12751 &  13.92710  & 1.72 & 1.94  &  3.09  \\
    \bottomrule
    \tabnotetext{a}{IRAC sources coincident with the radio-continnum peak(s) of \uchii regions with extended emission (\uchiie).}
  \end{tabular}
\end{table*}

Churchwell et al. (2006) using GLIMPSE images, identified and studied ``bubbles´´ around OB stars in the galaxy. Three-quarters of the bubbles in their sample are due to B4--B9 stars (too cool to produce detectable radio \hii regions), while the other ones are produced by young O--B3 stars with detectable radio H~II regions.  They suggest that bubbles overlapping known \hii regions appear to be produced by stellar winds and radiation pressure from young OB stars in massive star formation regions. Furthermore, they found that the 8.0~$\mu$m emission is weak at the center, but strong in the shells that define the bubbles and often extends outside the bubbles (see Figure 1). They support the idea that PAHs are destroyed in the vicinity of hot stars and instead trace the Photodissociation regions (PDRs) in the neighborhood of hot stars or star-clusters.

In our sample, the \uchii regions with EE seem to be part of larger structures (parsecs) that by size could be classified as classical \hii regions (parsecs--scale) and could be the ``bubbles´´ identified by Churchwell et al. (2006). Indeed our regions ressemble the morphologies described by these authors (contrast their Figure 2 with our Figures 1 to 3), and the cometary arcs observed at RC could be related with PDRs. 

\begin{figure}[!t] 
  \includegraphics[width=\columnwidth]{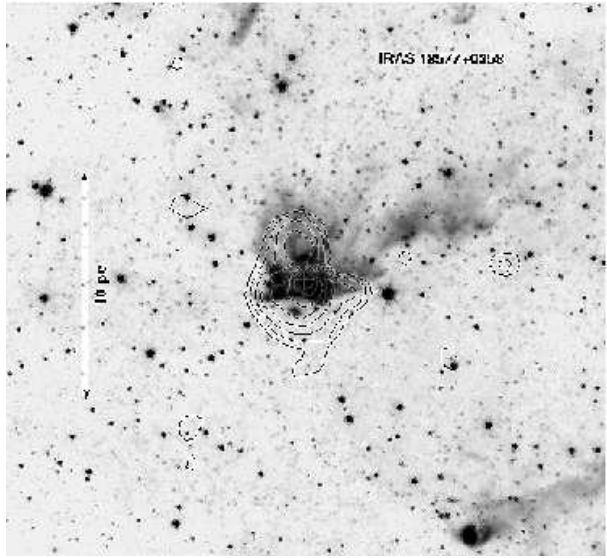}
  \caption{Gray scale IRAC 4--bands combined image of the \uchii region IRAS 18557+0358 (G37.55--0.11). Contours tracing the EE are from the RC VLA conf. D map (F07). The cross show the position of the \uchii region.}
  \label{fig:simple}
\end{figure}

In the color images, black filaments due to a higher density gas/dust component of IR dark clouds (IRDCs; e.g., Rathborne et al. 2006 and references therein) are observed. Because the IRDCs appear in silhouette against the bright Galactic background at mid-IR wavelengths, in Figures 1 to 3 these objects appear as white filaments. There is not a good correlation between the IR and RC morphology of IRDC's, nevertheless IRDC's are related with massive star formation regions, and several of them (e.g. IRAS 18222+1321) clearly cross through the center of the \uchii regions.

Images show the presence of dark patches in several \uchiie, which we suspect are IRDCs. It has been suggested that IRDCs could be the initial stage of massive star formation (similar to the pre-stellar cores of low mass star formation) and star clusters (e.g., Rathborne et al. 2006). If some IRDCs are forming high-mass stars and star clusters, and if the massive stars form in clusters with many lower-mass stars, then our imagery favors the idea that the IRDCs are related with the initial stages of both high and low--mass star formation, although because IRDCs are distant, the observations have so far lacked the sensitivity to detect lower-mass protostars.

\section{IRAC Photometry}
\label{sec:photo}

From the list of \uchiie with available IRAC observations presented in Table~4, 
a sub-sample of 13 \uchiie was selected to be studied in more detail. These regions lie on the galactic plane in the interval 20\arcdeg$\lesssim$~{\bf \emph{l}}~$\lesssim$50\arcdeg of galactic longitude.

RC maps of this sub-sample show --in some cases-- more than one peak. Thus we identify 19 \uchiis within these 13 regions.

IRAC photometry available on-line\footnote{\emph{http://irsa.ipac.caltech.edu/applications/Gator/} and then follow links to the highly reliable GLIMPSE~I and GLIMPSE~II catalogs of the \emph{Spitzer Space Telescope} Legacy Science Programs.} for these 13 regions was downloaded, searching for sources within a radius of 5\arcmin ~ around the EE and embracing the \uchii source(s). Although the photometry of a number of IR sources is obtained (see Table 3),
no data were available for the IR counterpart of \uchii source(s), found --by visual inspection-- to be within a tolerance separation of $\sim$ 3\arcsec ~ from the radio-continnum emission peaks (see Figures 1 to 3).

Following the standard procedure\footnote{described at \emph{http://ssc.spitzer.caltech.edu/irac/iracphot.} \emph{html}} we used IRAF/qphot package on the 19 \uchiis to obtain individual point source photometry in all four IRAC bands.
This process was performed on the pipeline mosaic images with 1.22\arcsec~ of pixel size. The source radius and the sky annulus were typically chosen to be 5--7~pix and 5--10/10--20~pix respectively. The greatest sky annulus value was used for radius greater than 5~pix and to avoid the flux contribution by nearby sources.
Zero points were set to 17.30, 16.82, 16.33 and 15.69, while photons per ADU (epadu  IRAF/qphot keyword) were set to 3.3, 3.7, 3.8, and 3.8, both for IRAC bands (channels) 1 to 4, respectively. Since this measurements were made source by source, only in some of them an aperture correction was necessary to apply. Photometrical errors are estimated to be$\sim$0.2~mag. A summary of IRAC colors produced by these measurements is provided in Table 2.

\subsection{IRAC color-color diagram}
\label{ssec:ccd}

\begin{figure*}[!t]
\centering
  \includegraphics[bb=90 50 555 520,width=0.75\textwidth]{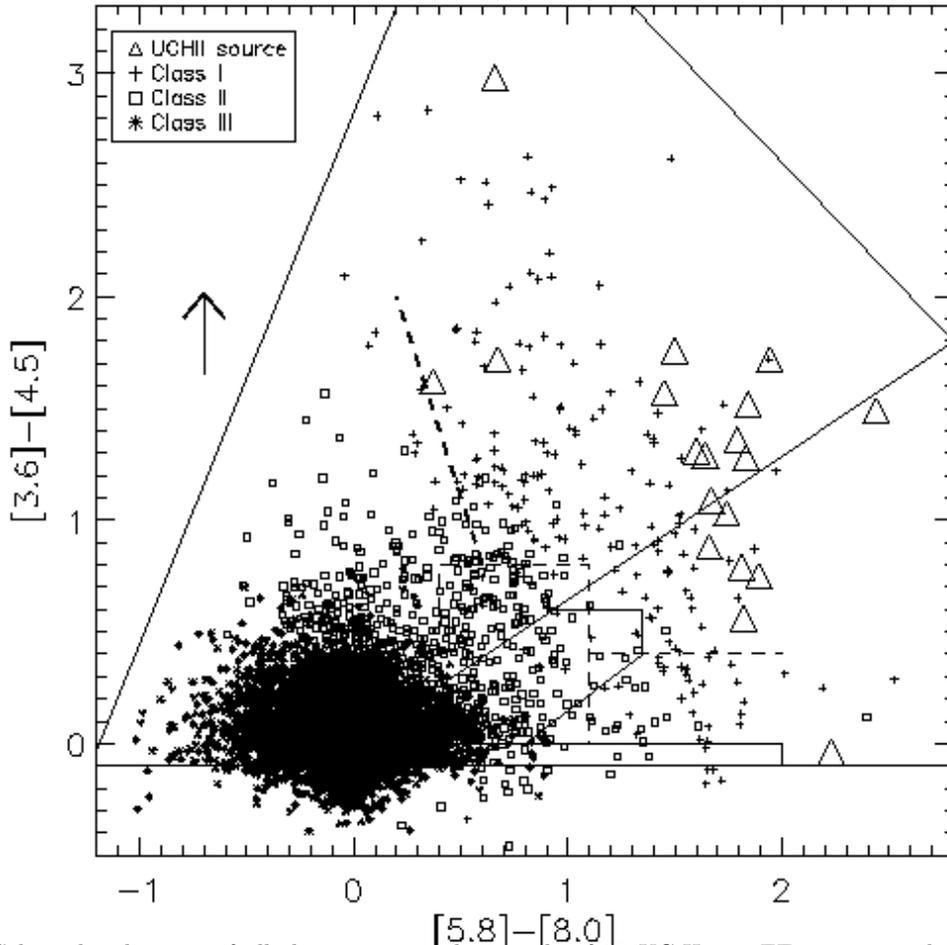}
  \caption{Color-color diagram of all the sources in the sample of 13 \uchiie regions, and with complete IRAC photometry. Dashed lines show the locus of Class~II sources as defined by Allen et al. 2004, the two lines outwards from this square mark the lower limit of the locus of Class~I sources, the upper one also shows the direction of interstellar extinction (Megeath et al. 2004, and references therein). Continous lines show the Stages I, II, III, and the region where Stages meet, defined by Rea06 based on a bast grid of theoretical models. Reddening vector of A$_V$=20 mag is also shown (Indebetouw et al. 2005).}
  \label{fig:ccd}
\end{figure*}

Once we completed the photometry for the \uchii sources whitin the EE regions we can plot both, these sources and the ones lying into the extended emission area obtained from GLIMPSE catalogs.
Figure 4 shows the [3.6]--[4.5] vs. [5.8]--[8.0] diagram of the 13 regions. Different symbols are used to mark the nature of the sources in Classes I, II, and III, following the classification scheme introduced by Lada (1987), but applied to IRAC wavelenghts. In this case, the SED slope ($\alpha_{IRAC}=d~log(\lambda F_\lambda) / d~log(\lambda)$) is calculated from 3.6 to 8.0~$\mu$m, and the classification criteria are: Class~I for sources with $\alpha > 0$, Class~II for $-2 \le \alpha \le 0$, and Class~III for $\alpha < -2$ (Lada et al. 2006, Chavarria et al. 2008). 
Although this classification was initially done for low/intermediate-mass young sources, an extension to the high-mass (up to 50 M$_\odot$) regime has been theoretically studied by Robitaille et al. (2006, Rea06 hereafter) and a similar definition of Stages I, II, and III has been introduced. One of the goals of our study is to define the observational locus of high-mass pre-main sequence stars into IRAC diagnostic diagramas. Thus, we have estimated $\alpha_{IRAC}$ also for the \uchii sources, their values are given in Table 2, and according to the previous classification, all except one (\uchii~17), would be Class~I.

Note that in Figure 4 there is a zone where $\sim$~75\% of the ultra-compact sources (denoted with triangles) are grouped, that locus is around \hbox{[5.8]--[8.0]$\simeq$1.7} and \hbox{0.5~$\lesssim$~[3.6]--[4.5]~$\lesssim$~2.0}. They are lying towards the upper-right extreme of the Stage~I sources and downwards into the region where all Stages~I, II, and III can meet (see, Figure~23 in Rea06). And this locus spans more or less in the direction of extinction vectors.

\subsection{AM product diagram}
\label{ssec:amd}

\begin{figure*}[!t]
\centering
  \includegraphics[bb=70 50 520 445,width=0.8\textwidth]{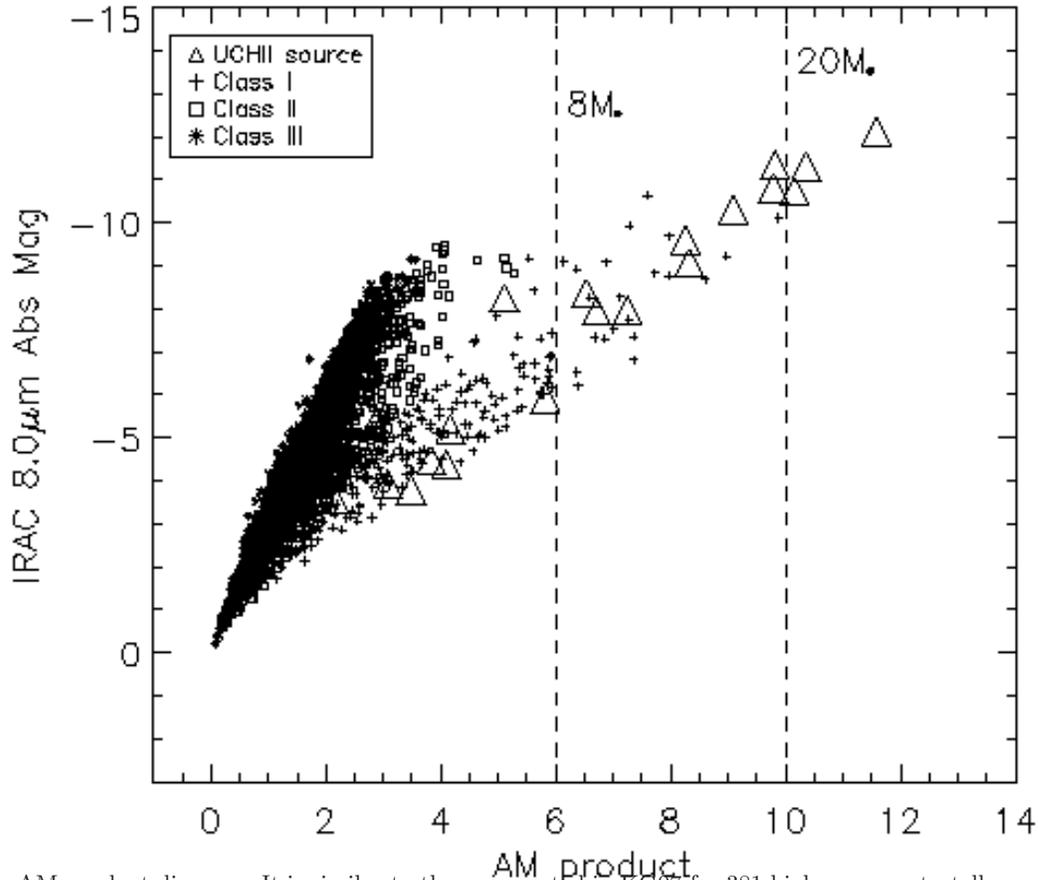}
  \caption{AM-product diagram. It is similar to those presented in KG07 for 381 high-mass protostellar candidates regions with GLIMPSE data available. Note that in this diagram, \uchii sources are lying on the edge of the ``extreme'' Class~I sources.}
  \label{fig:am8}
\end{figure*}

Following KG07, a definition of the 8.0~$\mu$m magnitude and $\alpha_{IRAC}$ product is given by $AM = -M_{8\mu m}\times(\alpha_{IRAC} + 6)/10$, and is called ``AM product'' (alpha-magnitude product). This definition, and then a plot of absolute magnitude in 8~$\mu$m vs. AM product, help to separate effectivelly the luminous sources and to compare between sources at different distances.
Photometric 8~$\mu$m magnitudes of this sub-sample are between $\sim$1.5--7~mag, and distances between 3.5 and 13.5 kpc. A plot of IRAC sources is shown in Figure 5. Vertical lines at AM product values of 6 and 10, mark the approximate position of theoretical models by Rea06 for Class~I objects at 8 and 20~M$_\odot$, respectively (see, Fig.~3 in KG07).

A linear fit to the \uchiis, and Classes~I, II, and III, give the slopes -0.96, -0.97, -2.02, and -2.71, respectively. This values are close to 1, 2, and 3, as might be expected from the definition of AM product,
for Classes~I, II, and III, respectively. Negative values of the slopes come from the nature of the absolute magnitude values in $y$ axis. The small difference in slope between \uchiis and other Class~I sources can be seen in Figure 5 with the triangles lying on the edge of the ``extreme'' Class~I sources. This AM product plot suggests that some of the \uchiis might be Class~0 candidates.

\subsection{IRAC sources by class}
\label{ssec:stats}

The total number of sources by class in the sample of 13 \uchiie regions are: 8172 Class~III, 902 Class~II, 196 Class~I, and 19 \uchiis. Source counts in each region by Class are listed in Table 3. Relative percentages to the number of IR sources by Class in each region are given in parenthesis. 

\begin{table}[!t]\centering
\small
  \setlength{\tabnotewidth}{\columnwidth}
  \tablecols{5}
  \setlength{\tabcolsep}{0.5\tabcolsep}
  \caption{Counts of IRAC sources\tabnotemark{a}} \label{tab:counts}
  \begin{tabular}{rrrrr}
    \toprule
   \multicolumn{1}{c}{IRAS name} & \multicolumn{1}{c}{N\tabnotemark{b}} & \multicolumn{1}{c}{Class I\tabnotemark{c}} & \multicolumn{1}{c}{Class II\tabnotemark{c}} & \multicolumn{1}{c}{Class III\tabnotemark{c}} \\
    \midrule
   18097--1825A& 1100\,\,\,\,\,\,\,\, & 19\,(2)&  80\,\, (7) &1001\,(91)\\
   18222--1321\,\,\,\,\, &  531\,\,\,\,\,\,\,\, & 18\,(3)& 101\,(19) & 412\,(78)\\
   18248--1158\,\,\,\,\, & 1028\,(4) & 21\,(2)&  57\,\, (6) & 950\,(92)\\
   18311--0809\,\,\,\,\, &  846\,(2) &  7\,(1)& 100\,(12) & 739\,(87)\\
   18353--0628\,\,\,\,\, &  810\,(2) &  7\,(1)&  80\,(10) & 723\,(89)\\
   18402--0417\,\,\,\,\, &  743\,\,\,\,\,\,\,\, & 17\,(2)& 117\,(16) & 609\,(82)\\
   18469--0132\,\,\,\,\, &  724\,\,\,\,\,\,\,\, & 10\,(1)&  47\,\, (6) & 667\,(92)\\
   18538+0216\,\,\,\,\,  &  681\,\,\,\,\,\,\,\, & 24\,(4)&  78\,(11) & 579\,(85)\\
   18577+0358\,\,\,\,\,  &  535\,\,\,\,\,\,\,\, & 10\,(2)&  62\,(12) & 463\,(86)\\
   18593+0408\,\,\,\,\,  &  583\,\,\,\,\,\,\,\, & 12\,(2)&  63\,(11) & 508\,(87)\\
   19120+0917\,\,\,\,\,  &  858\,\,\,\,\,\,\,\, &  9\,(1)&  34\,\, (4) & 815\,(95)\\
   19120+1103\,\,\,\,\,  &  437\,(2) & 26\,(6)&  50\,(11) & 361\,(83)\\
   19181+1349\,\,\,\,\,  &  394\,\,\,\,\,\,\,\, & 16\,(4)&  33\,\, (8) & 345\,(88)\\
    \bottomrule
    \tabnotetext{a}{Without including \uchii sources, but see note\tabnotemark{b}}
    \tabnotetext{b}{Number of sources with all four IRAC-bands data within a 5$\arcmin$ radius. In parenthesis, the number of \uchii sources is given, when there are more than one in the region.}
    \tabnotetext{c}{Numbers in parenthesis show the relative percentage.}
  \end{tabular}
\end{table}

Although the range of these sources varies form about 400 to 1000, the percentages are more or less consistent from region to region, and in average, $\sim$88\% are Class~III sources, $\sim$10\% are Class~II, and $\sim$2\% are Class~I. These regions where selected because they embrace at least one young massive star evidenciated by the radio-continnum peaks at 3.6~cm emission. And, it is interesting to note that the  percentage averages for regions with more than one \uchii source (IRAS~18248--1158, IRAS~18311--0809, IRAS~18353--0628, and IRAS~19120+1103) keep roughly the same values.

This numbers give an statistical base for a further study of the stellar evolutionary stage in these \uchiie regions, and its relation with the observed stellar mass function. The photometry of the larger sample of 22 \uchiie (listed in Table 1) will provide a more robust conclusion on these counts.

\section{Conclusions}
\label{sec:fin}
Because this is a work in progress, we list our preliminary conclusions.

\begin{enumerate}

\item Based on a comparison between RC maps at 3.6~cm (VLA conf.~D) and GLIMPS--IRAC images of a sample of \uchiie, a coincidence of morphologies is observed. This indicate that heated dust coexists with ionized gas in the EE areas. 
\item The peaks of RC emission are in most cases coincident with luminous IR counterparts. We
identify broad (FWHM~$\sim$8\arcsec) IR point sources in the \uchii position in several cases.
\item We perform \emph{Spitzer}--IRAC photometry of 19 \uchii sources in 13 \uchiie regions (a subsample of F07), that are not available at GLIMPSE photometry catalogs.
\item This 19 \uchiis lie in a region located (locus) at [5.8]--[8.0]$\simeq$1.7 and 0.5$\lesssim$[3.6]--[4.5]$\lesssim$2.0, in the ``classic'' IRAC color-color diagram.
\item UC~\hii sources also lie at the extreme border of Class~I objects in the AM product diagram (KG07), suggesting that at least some of them might be Class~0 candidates.
\item Following an $\alpha_{IRAC}$ classification, we find that in average, the population of sources are: $\sim$2\% of Class~I (proto-stars), $\sim$10\% of Class~II (stars with disks), and $\sim$88\% of Class~III (stellar photospheres), in \uchiie regions.
\end{enumerate}

\section{Acknowledgements}

A. P. would like to finish this proceeding with the congratulation frase given to Dr. Carrasco at the end of her talk: ``HAPPY 60's DR. CARRASCO --Thanks to teach me Astronomy and contribute to my formation as a researcher''.

E. d. l. F. acknowledges support from CONACyT (Mexico) grant 124449 and CONACyT--SNI exp. 1326. 

This work is based in part on observations made with the \emph{Spitzer Space Telescope}, which is operated by the Jet Propulsion Laboratory, California Institute of Technology under a contract with NASA.

\end{document}